\documentclass[aps,showpacs,preprintnumbers,amsmath,amssymb]{revtex4}

\UseRawInputEncoding
\usepackage{graphicx}
\usepackage{epstopdf}
\usepackage{color}
\usepackage{subfig}
\usepackage{aasmacros}
\usepackage{amsmath,amssymb,amsfonts}

\begin{document}
    \captionsetup{justification=raggedright,singlelinecheck=false}

    \baselineskip=0.8cm
    \title{\bf Image of Bonnor black dihole with a thin accretion disk and its polarization information}

    \author{Zelin Zhang$^{1}$,
    Songbai Chen$^{1,2}$\footnote{Corresponding author: csb3752@hunnu.edu.cn},
    Jiliang Jing$^{1,2}$ \footnote{jljing@hunnu.edu.cn}}
    \affiliation{$^1$Department of Physics, Key Laboratory of Low Dimensional Quantum Structures
    and Quantum Control of Ministry of Education, Synergetic Innovation Center for Quantum Effects and Applications, Hunan
    Normal University,  Changsha, Hunan 410081, People's Republic of China
    \\
    $ ^2$Center for Gravitation and Cosmology, College of Physical Science and Technology, Yangzhou University, Yangzhou 225009, People's Republic of China}

    \begin{abstract}
    \baselineskip=0.6 cm
    \begin{center}
    {\bf Abstract}
    \end{center}
    We have studied the image of Bonnor black dihole surrounded by a thin accretion disk where the electromagnetic emission is assumed to be dominated respectively by black body radiation and synchrotron radiation.
    Our results show that the intensity of Bonnor black dihole image increases with the magnetic parameter and the inclination angle in both radiation models. The image of Bonnor black dihole in the synchrotron radiation model is one order of magnitude brighter than that in the black body radiation model, but its intensity in the former decreases more rapidly with the radial coordinate. Especially, for the synchrotron radiation model, the intensity of the secondary image is stronger than that of the direct image at certain an inclination angle. We also present the polarization patterns for the images of Bonnor black dihole arising from the synchrotron radiation, which depend sharply on the magnetic parameter and inclination angle. Finally, we make a comparison between the polarimetric images of Bonnor black dihole and M87*. Our result further confirms that the image of black hole depends on the black hole's properties itself, the matter around black hole and the corresponding radiation occurred in the accretion disk.

    \end{abstract}

    \maketitle
    \newpage

    \section{Introduction}

    It is well known that the observational black hole astronomy has been entered an exciting era of rapid progress due to the detection of gravitational waves \cite{2016PhRvL.116f1102A,2016PhRvL.116x1103A,2017PhRvL.118v1101A,2017PhRvL.119n1101A,2017ApJ...851L..35A} and the releasing of the images of the black hole M87* \cite{2019ApJ...875L...1E,2019ApJ...875L...2E,2019ApJ...875L...3E,2019ApJ...875L...4E,2019ApJ...875L...5E, 2019ApJ...875L...6E} and Sgr A* \cite{2022APJ...930L...17E}. Recently, the Event Horizon Telescope (EHT) collaboration have released the first polarized images of the black hole M87* \cite{2021ApJ...910L..12E,2021ApJ...910L..13E}, which provides a new way to probe the matter distribution, the electromagnetic interaction and the accretion process around black hole. It is because the brightness of the surrounding emission region and the polarization pattern in the image carry a wealth of information about the electromagnetic emissions near the black hole. Thus, a lot of effort has been devoted to making theoretical research on the polarized images of black holes \cite{2021ApJ...912...35N,2021PhRvD.104d4060G}. Recently, with a simple ring model, the polarized images of axisymmetric fluid orbiting arising from synchrotron emission in various magnetic field have been investigated for Schwarzschild black hole \cite{2021ApJ...912...35N} and Kerr black hole \cite{2021PhRvD.104d4060G}. Although only the emission within a narrow range of radii $R$ is considered in this simple model, it clearly reveals that the polarization signatures are dominated by magnetic field configuration, black hole spin and observer inclination. With this model, the polarized image of an equatorial emitting ring has been studied for 4D Gauss-Bonnet black hole \cite{2021arXiv211110138Q} and regular black holes \cite{2022arXiv220500391L}. The polarized distribution of the image originating from synchrotron radiations of the charged point particle in curved spacetime has been investigated in \cite{2022arXiv220504777Z,2022arXiv220302908H}. Moreover, the polarized image of a Schwarzschild black hole with a thin accretion disk has been researched under the interaction between electromagnetic field and Weyl tensor \cite{2021EPJC...81..991Z}.

    Nowadays, several detected gravitational waves events are caused by binary black hole merger \cite{2016PhRvL.116f1102A,2016PhRvL.116x1103A,2017PhRvL.118v1101A,2017PhRvL.119n1101A,2017ApJ...851L..35A} or by binary neutron star merger \cite{LIGOScientific:2017fdd,LIGOScientific:2017zic}, which means that the binary black hole systems should be common in astrophysical systems. The shadow of some binary black hole systems have been investigated, which indicates that their shadows own special features differed from that of an isolated single black hole. D. Nittathe studied the shadow of the colliding of the two black hole with a Kastor-Traschen solution \cite{Nitta:2011in}. It is shown that as two black hole stand too close, there exists the eyebrow-like shadows appeared around the main shadows, which is caused by the photon's chaotic lensing. The similar features also appear in other binary black holes cases including Majumdar-Papapetrou binary black holes \cite{Bohn:2014xxa}, binary black holes inspiral space-time \cite{Shipley:2016omi}, double-Schwarzschild and double-Kerr black holes \cite{Cunha:2018cof}, and Bonnor black dihole \cite{2018PhRvD..97f4029W}. These special features of shadows could provide a new antenna to probe binary black hole system.

    The above investigations of binary black hole shadows are based on the assumption that the light source is distributed homogeneously in the total celestial sphere. However, in the real astronomical environment, the actual light source should be accretion disk around black hole, which must yield that the black hole image differs from that casted by the former homogeneous light source. In 1969, the accretion disk was proposed to explain the huge energy released by quasars \cite{1969Natur.223..690L}. The current models of accretion disks around black holes can be generally classified into the cold disk model and the hot one. The standard accretion disk model \cite{1973A&A....24..337S} belongs to the cold disk model where
    the disk owns high conversion efficiency and low temperature. The gravitational potential energy of the fluid particle in the mass accretion process can be rapidly converted into the outgoing radiation. Moreover, the disk has the relative high surface density and the large optical depth so that the radiation emitted by the disk surface can be considered as a perfect black body radiation. In the hot disk model, the disk has high temperature and the radiation emitted by the disk is dominated by synchrotron radiation. These disk models have been widely used to explain various phenomenon occurred in the accretion disks around black holes \cite{1973blho.conf..343N, 1974ApJ...191..499P,2013LRR....16....1A, 2022JCAP...02..034L,2022PhRvD.105b3021B,2021arXiv211201747G,2021PhRvD.104j4055S,2021arXiv211009820R,2021EPJC...81..840Z,2021PhRvD.104d4049G,2021EPJC...81..473H,2021ApJ...910...52C,2020PhRvD.101b3002F,2019arXiv191013259C} . The shadows of black holes with a thin disk have been studied in \cite{1979A&A....75..228L,2018Univ....4...86L,2021EPJC...81..885G,2022IJMPD..3150041L}.

    As in the previous discussion, the current study of the image of binary black hole system mainly focus on the cases without accretion disk. In this paper, we will study the image of Bonnor black dihole \cite{1966ZPhy..190..444B} surrounded by a light thin accretion disk. The metric of Bonnor black dihole describes the geometry of a compact object with a dipole magnetic field. Its two magnetic poles with different signs are located on the axis of symmetry. The special structures of the Bonnor spacetime from the dipole magnetic field make the dynamical system of particle non-integrable so that there exists chaos in the the motion of particles \cite{2012AAS...22043007K,2014SPPhy.157..373K,2015arXiv150404844A}. In this paper, we will firstly focus on the case where the electromagnetic emission in the disk is assumed to be dominated by black body radiation and then the disk is treated as the source of natural light. Then, we also consider the other case where the electromagnetic emission in the disk is dominated by synchrotron radiation and the light emitted by disk is the linear polarization light. We also further study the image of Bonnor black dihole and its polarization vector distribution in this case. Since Bonnor black dihole carries magnetic field itself, it will affect the polarization of light arising from synchrotron radiation in the disk. Thus, in this paper, we will probe the effect of magnetic dipole parameter on the image of Bonnor black dihole surrounded by a thin accretion disk and on its corresponding polarized patterns.

    This paper is organized as follows: In section \ref{sec2}, we will briefly review the Bonnor black dihole spacetime and its magnetic field distribution. In section \ref{sec3}, we apply ARCMANCER \cite{2018ApJ...863....8P} to simulate the image of Bonnor black dihole surrounded by a light thin accretion disk which is dominated by black body radiation. In section \ref{sec4}, we also present the image of Bonnor black dihole surrounded by the thin accretion disk which is dominated by synchrotron radiation and present the polarization image of Bonnor black dihole. Finally, we conclude this article with a summary.

    \section{SPACETIME OF BONNOR BLACK DIHOLE}
    \label{sec2}
    Let us now to review briefly the solution of Bonnor black dihole which was obtained by Bonnor \cite{1966ZPhy..190..444B}. The solution of dihole describes a static massive source with a magnetic dipole in Einstein-Maxwell theory, and its metric form can be expressed as \cite{1966ZPhy..190..444B}
    \begin{equation}
        \label{metric}
        ds^{2}=-\left(\frac{P}{Y}\right)^2 dt^2+\frac{P^2 Y^2}{Q^3 Z}\left(dr^2 + Z d \theta^2\right)+\frac{Y^2 Z \sin ^2 \theta}{P^2} d \phi ^2 ,
    \end{equation}
    with
    \begin{eqnarray}
        P&=&r^{2}-2mr-b^{2}\cos^{2}\theta,\quad\quad\quad Q=(r-m)^{2}-(m^{2}+b^{2})\cos^{2}\theta,\\
        Y&=&r^{2}-b^{2}\cos^{2}\theta,\quad\quad\quad \quad\quad\quad Z=r^{2}-2mr-b^{2}.
    \end{eqnarray}
    The total mass of Bonnor black dihole $M$ and the magnetic dipole moment $\mu$ are related to two distinct independent parameters $m$ and $b$ by $ M = 2m$ and $\mu = 2 m b$, respectively.
    It is easy to see that this spacetime is asymptotically flat since the metric reduces to the Minkowski case  as the polar coordinate $r$ approaches to infinity. However, as the parameter $b=0$, this metric (\ref{metric}) tends tothe Zipoy-Voorhees metric with $\delta= 2$ rather than  Schwarzschild one, which describes a monopole of mass $2m$ together with higher mass multipoles depended on the parameter $m$. Moreover, from the null hypersurface condition
    \begin{eqnarray}
        g^{\mu\nu}\frac{\partial f}{\partial x^{\mu}}\frac{\partial f}{\partial x^{\nu}}=0,
    \end{eqnarray}
    one can find that the event horizon of the spacetime (\ref{metric}) lies at $r_h=m+\sqrt{m^2+b^2}$. Especially, the region outside the horizon is regular since $g_{\phi\phi}>0$ in this region and there exist no any closed timelike curves outside the horizon.

    The electromagnetic four vector potential $A_{\mu}$ of Bonnor space-time~(\ref{metric}) is given by
    \begin{equation}\label{diancichang1}
        A_{\mu} = (0,0,0,\frac{2mbr\sin^{2}\theta}{P}),
    \end{equation}
    where $\mu = 0,1,2,3$ correspond to the coordinates $(t,r,\theta,\phi)$, respectively. The electromagnetic tensor $F_{\mu\nu}=A_{\nu;\mu}-A_{\mu;\nu}$ can be expressed as
    \begin{equation}
        \label{Fmunu}
        F_{\mu\nu} = \left(
            \begin{array}{cccc}
             0 & 0 & 0 & 0 \vspace{1ex} \\
             0 & 0 & 0 & -\frac{2 m b \sin^2\theta \left(r^2+ b^2 \cos ^2\theta\right)}{P^2} \vspace{1ex} \\
             0 & 0 & 0 & \frac{2 m b r Z \sin 2 \theta }{P^2} \vspace{1ex} \\
             0 & \frac{2 m b \sin ^2\theta \left(r^2 + b^2 \cos ^2\theta \right)}{P^2} & -\frac{2 m b r Z \sin 2 \theta }{P^2} & 0 \vspace{1ex} \\
            \end{array}
            \right).
    \end{equation}
    The electromagnetic field in the spacetime (\ref{metric}) acts as a matter source in the Einstein field equations and the non-zero components in the energy-momentum tensor $T^{\mu\nu}$ are
    \begin{eqnarray}
        \label{emt0}
        T^{00}&=&-\frac{2 b^2 m^2 Q^3 \sin ^2\theta \left(\left(b^2 \cos ^2\theta+r^2\right)^2+4 r^2 Z \cot ^2\theta\right)}{P^6 Y^2}, \\
        T^{11}&=&-\frac{2 b^2 m^2 Q^6 Z \sin ^2\theta \left(\left(b^2 \cos ^2\theta+r^2\right)^2-4 r^2 Z \cot ^2\theta\right)}{P^6
        Y^6}, \\
        T^{22}&=&\frac{2 b^2 m^2 Q^6 \sin ^2\theta \left(\left(b^2 \cos ^2\theta+r^2\right)^2-4 r^2 Z \cot ^2\theta\right)}{P^6 Y^6},\\
        T^{12}&=&T^{21}=\frac{4 b^2 m^2 Q^6 r Z \sin 2 \theta \left(b^2 \cos ^2\theta+r^2\right)}{P^6 Y^6}, \\
        T^{33}&=&-\frac{2 b^2 m^2 Q^3 \left(\left(b^2 \cos ^2\theta+r^2\right)^2+4 r^2 Z \cot ^2\theta\right)}{P^2 Y^6 Z}.\label{emt1}
    \end{eqnarray}
Clearly, the electromagnetic field (\ref{diancichang1}) affects the synchrotron radiation in the disk and then changes the polarization properties of the image of the black dihole.

    \section{IMAGE OF BONNOR BLACK DIHOLE WITH A THIN ACCRETION DISK dominated by black body radiation}
    \label{sec3}

    In this section, we will consider Bonnor black dihole system with a cold thin accretion disk \cite{1974ApJ...191..499P} in the equation plane. The fluid particles in the disk are assumed to move along geodesic circular orbits, which means that the specific energy $E$, the specific angular momentum $L$, and the angular velocity $\Omega$ of the fluid particles have the form
    \begin{eqnarray}\label{elo}
        E=-\frac{g_{tt}}{\sqrt{-g_{tt}-g_{\phi\phi}\Omega^2}}, \quad\quad
        L= \frac{g_{\phi\phi} \Omega}{\sqrt{-g_{tt}-g_{\phi\phi}\Omega^{2}}}, \quad\quad
        \Omega=\sqrt{-\frac{g_{tt,r}}{g_{\phi\phi,r}}}.
    \end{eqnarray}
    In the cold disk model, the gravitational potential energy is released eventually mainly in the form of electromagnetic radiation and the radiation emitted by the disk surface can be considered as a perfect black body radiation. The degree of polarization of electromagnetic radiation is lower in this model. For simplification, we here assume that the thin disk around the Bonnor black dihole is modeled by a steady state accretion disk \cite{1974ApJ...191..499P} and the time-averaged energy flux emitted from the disk's surface can be expressed as \cite{1974ApJ...191..499P}
    \begin{equation}\label{eqfs}
    \mathcal{F}_{s}(r) = -\frac{\dot{M_0}}{4\pi\sqrt{-G}}\frac{\Omega_{,r}}{\left(E-\Omega L\right)^2}\int_{r_{mb}}^{r}(E-\Omega L)L_{,r} \,dr,
   \end{equation}
   where $\sqrt{-G}=\frac{r^3(r-2m)}{(r-m)^3}$ for the Bonnor black dihole spacetime. $M_0$ is the accretion rate and $r_{mb}$ is the innermost stable circular orbit (ISCO) for the inner edge of the disk. For the sake of simplicity, here we do not consider the viscosity of the disk.  The effect of the magnetic field in the spacetime (\ref{metric}) on the thin disk is considered through the quantities $\Omega$, $E$, $L$, $G$ and the radius of ISCO $r_{mb}$. By means of the invariant intensity $I =\mathcal{I}_{\nu}/\nu^3$, the specific intensity observed by a
distant observer can be related to the emitted specific intensity by $\mathcal{I}_o=g^3 \mathcal{I}_s$ and the redshift factor $g=1/(1+z) = {\nu_o}/{\nu_s}$. The radiation intensity $\mathcal{I}_s$ can be obtained by $\mathcal{I}_s= d\mathcal{F}_{s}/ (\cos{\theta} d\Omega$). The photon emitting from the disk moves along null geodesic to the observer in the far region. With these formulas, we apply the code ARCMANCER \cite{2018ApJ...863....8P} to simulate numerically the image for the Bonnor black dihole and probe effects of the magnetic dipole on the image.
    \begin{figure}[htbp]
        \centering
        \includegraphics[width=16cm]{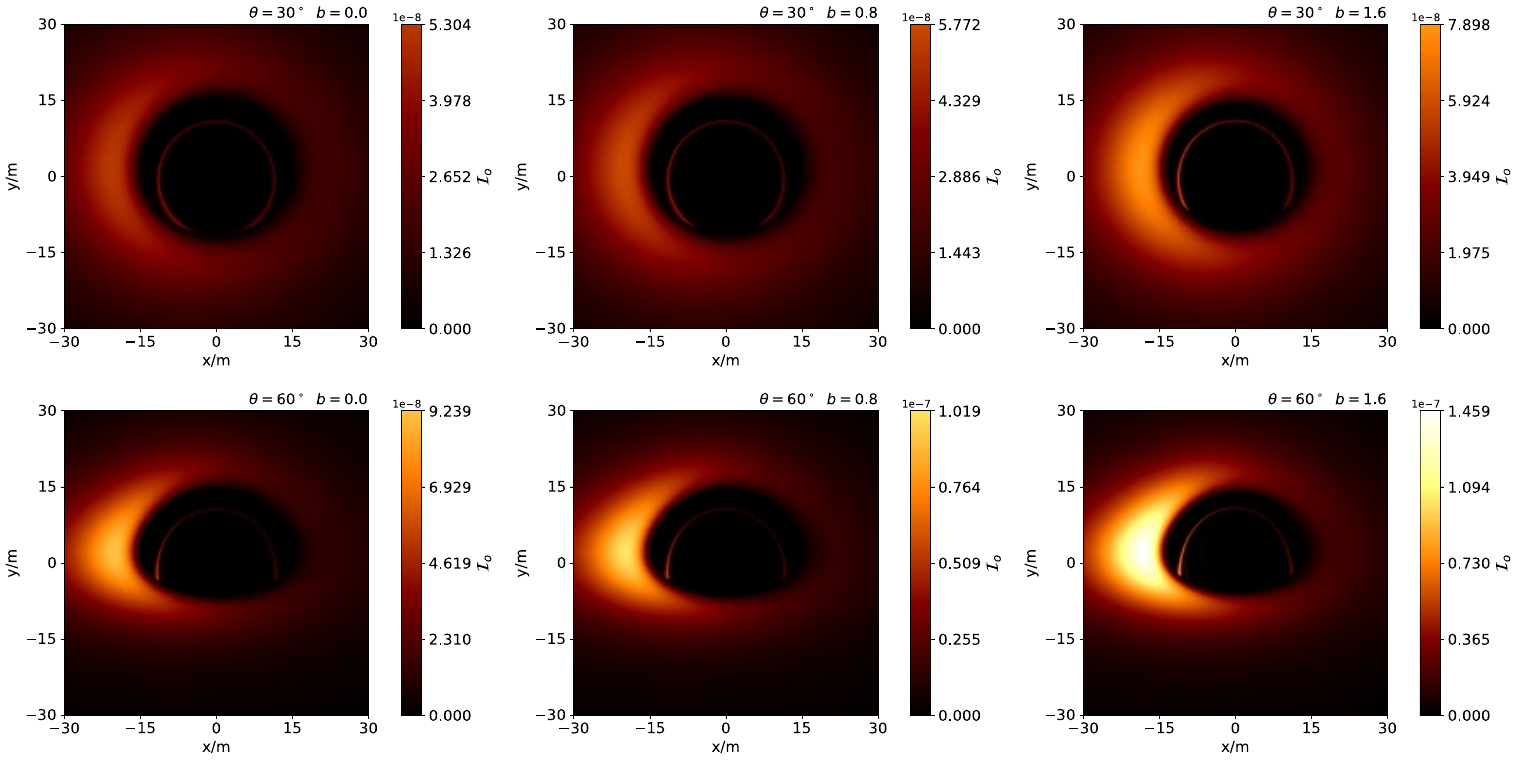}
        \caption{Image of Bonnor black dihole surrounded by a thin accretion disk with different magnetic dipole parameter $b$ for the fixed $r_0=1000$. The upper row is for the observer's inclination angles $\theta_0=30^{\circ}$ and the bottom row is for $\theta_0=60^{\circ}$ . In each row, the magnetic dipole parameter $b$ from left to right is taken to 0, 0.8 and 1.6 respectively. Here, we set $m=1$. }
        \label{f1}
    \end{figure}
    \begin{figure}
        \centering
        \includegraphics[width=12cm]{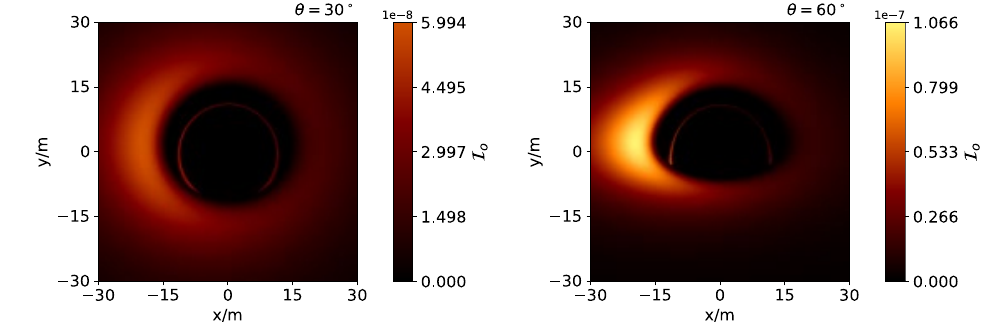}
        \caption{Image of Schwarzschild black hole surrounded by a thin accretion disk. The mass of Schwarzschild black hole here is set to the same as that of Bonnor black dihole in Fig.\ref{f1}.}
        \label{f3}
    \end{figure}

\begin{figure}[htbp]
    \centering
    \includegraphics[width=16cm]{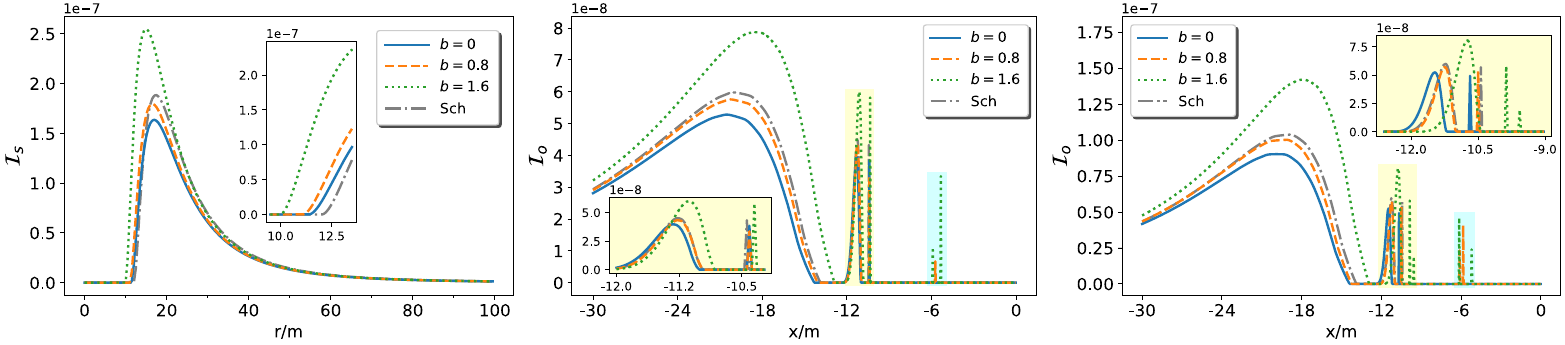}
    \caption{Intensity distribution curve with magnetic dipole parameter $b$ along the line $y=0$. The gray dashdot line in each plot corresponds to the Schwarzschild case. The left panel denotes the intensity of source $\mathcal{I}_s$ in the disk surface. The middle and the right correspond to the observed intensity $\mathcal{I}_o$ with inclination angle $\theta_0=30^{\circ}$ and $60^{\circ}$, respectively.}
    \label{f2}
    \end{figure}

    Fig.\ref{f1} shows the intensity of Bonnor black dihole image almost increases with the magnetic parameter $b$ and the observer's inclination angles $\theta_0$. The main reason of the image intensity increasing with $b$ is that intensity of source in the disk surface increases with the parameter $b$, which is shown in the left panel in Fig.\ref{f2}.
    Since we here assume that the accretion disk rotates counterclockwise, the image on the left is brighter than that on the right, which is caused by the so-called Doppler beaming effect. Moreover, we find that the intensity in the bright region of the image with the inclination angle $\theta_0=60^{\circ}$ is stronger than that with $\theta_0=30^{\circ}$, which can be explained by a fact that the observed intensity of image depends on the angle between the observer screen and the fluid particle's velocity in the disk. Here, for comparison, we also  present the image of Schwarzschild black hole with the same mass parameter and observer as in Fig.\ref{f3}. It is shown that the distribution of the radiation in the Schwarzschild case is similar to that in the Bonnor black dihole case. But the maximum intensity in the Schwarzschild case is stronger than that in the Bonnor case with $b=0$ and $b=0.8$, and is weaker than that in the case of $b=1.6$.
    Furthermore, we also present the intensity distribution curve of image along the line $y=0$, which confirms further that the intensity in the image of black dihole increases with $b$. This also means that the change of intensity in the image with the parameter $b$ is dominated by properties of the intensity of the source in the disk surface in this case. Although there is only a single peak in the intensity curve of source $\mathcal{I}_s$ for the fixed $b$, one can find that there are some peaks in the curve of intensity $\mathcal{I}_o$, which correspond to the extremal maximum of intensity in the direct image and the secondary image, respectively. This is caused by the strong gravitational lensing near the black dihole. Moreover, in the case $b=0.8$ and $b=1.6$, we also find some additional peaks marked in light blue that do not appear in the case of $b=0$ and Schwarzschild black hole. This may be fine structures in the image of Bonnor black dihole yielded by the chaotic lensing as discussed in \cite{2018PhRvD..97f4029W}.
    Fig.\ref{f2} also show that the intensity in the bright region of the image increases with the inclination angle, which is similar to that shown in Fig.\ref{f1}.
    \begin{figure}[htbp]
        \centering
        \includegraphics[width=16cm]{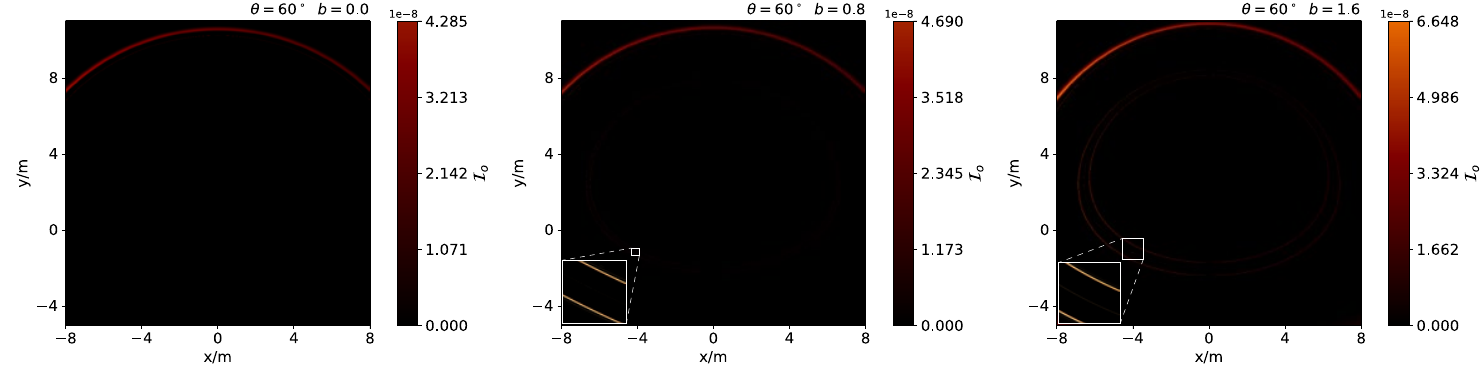}
        \caption{The fine structures in the image of Bonnor black dihole. In the middle and right panels, the subgraph in the lower left corner is a partially enlarged image of the region marked with a small white square.}
        \label{f4}
    \end{figure}
    \begin{figure}[htbp]
        \centering
        \includegraphics[width=16cm]{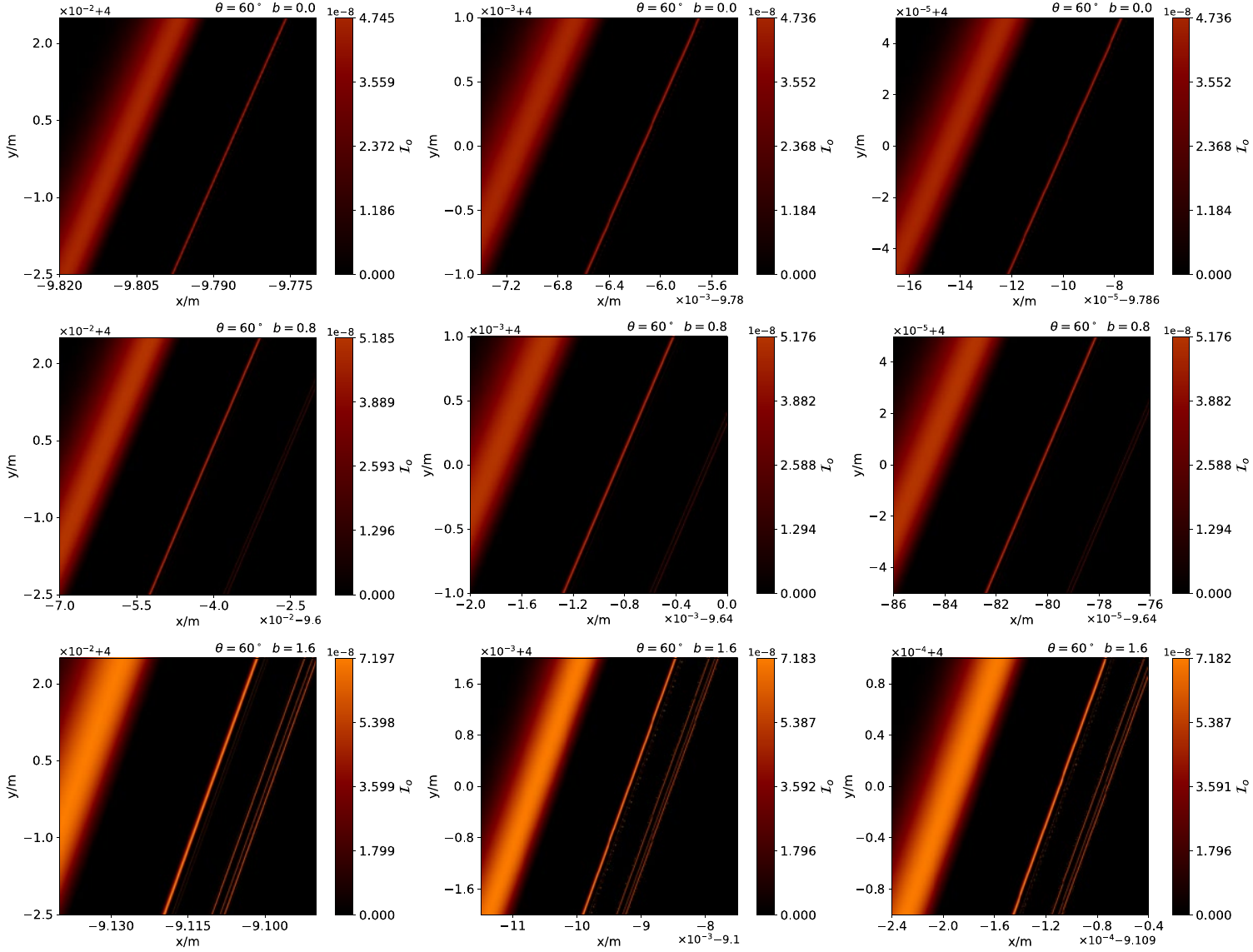}
        \caption{Enlarged image near secondary image with $\theta_0=60^{\circ}$ for different $b$. The panels from the top to bottom denote the cases with $b=0$, $0.8$ and $1.6$, respectively.}
        \label{f5}
    \end{figure}

    In Figs. \ref{f4} and \ref{f5}, we present higher order images and some fine structures in the image of Bonnor black dihole. As $b=0$, we find that there is no bright pattern in the black area in Figs. \ref{f4} and \ref{f5}. As $b=0.8$ or $1.6$, one can find an ``eyebrow shape" bright region appeared in the black area, which is similar to the result obtained in \cite{2018PhRvD..97f4029W}. However, the pattern of ``eyebrow shape" bright region is not quite the same as that in \cite{2018PhRvD..97f4029W}. The main reason is that here the light source is a thin disk with an inner boundary $r_{\rm in}=r_{\rm ISCO}$ lied in the equation plane, while in Ref.\cite{2018PhRvD..97f4029W} the light source is distributed homogeneously in the total celestial sphere. With the increase of $b$, one can find that the ``eyebrow shape" bright lines contained fine structures become more obvious. The appearance of ``eyebrow shape" images could be treated as a main feature of black hole image by photon chaotic lensing because the dynamical system of photon is non-integrable in the Bonnor black dihole background and then the chaotic motion of photons could appear.

    After enlarging the region near the secondary image, we presents rich fine structures as shown in Figs. \ref{f5}. As $b=0$, the fine structures near the secondary image actually are compose of much higher order images, which are caused by the photons passing through the disk several times before their reaching the observer. Moreover, each order indirect image has only a single image and the order of indirect images depends on only the number of times of photons passing through the disk in this case. However, as $b=0.8$ or $1.6$, there are some extra images near the above indirect images. These extra images own self-similar fractal structures, but they originate from the photon chaotic lensing near Bonnor black dihole since the number of these extra images increases with the magnetic parameter $b$. Therefore, the existence of magnetic moment in spacetime makes the motion of photons more complex and then brings richer fine structures in the image of Bonnor black dihole.

    \section{IMAGE OF BONNOR BLACK DIHOLE WITH A THIN ACCRETION DISK dominated by synchrotron radiation}
        \label{sec4}
    The synchrotron radiation is another important radiation occurred in the disk around black hole. Especially, the electromagnetic radiation caused by the synchrotron radiation owns much higher degree of polarization, which is helpful to understand the polarized image of black hole. Thus, in this section, we consider a thin accretion disk dominated by the synchrotron radiation in the equatorial plane around Bonnor black dihole.

    For a Bonnor dihole (\ref{metric}), it carries a magnetic field itself related to the electromagnetic tensor (\ref{Fmunu}), which plays an important role on the polarized image of the accretion disk around the dihole. In order to study the polarized image of dihole, we must first get the polarization vector of photon emitted by fluid in the disk \cite{2021ApJ...912...35N,2021PhRvD.104d4060G}. In the local Cartesian frame $\{e_{\hat{t}}, e_{\hat{r}}, e_{\hat{\theta}}, e_{\hat{\phi}} \} $ of the arbitrary point $P$ in the disk ($P$-frame ), the electromagnetic tensor $F_{\mu\nu}$ (\ref{Fmunu}) can be re-expressed as
    \begin{equation}
        F_{\hat{\alpha}\hat{\beta}} = F_{\mu\nu} e_{\hat{\alpha}}^{\mu} e_{\hat{\beta}}^{\nu},
    \end{equation}
    where $e_{\hat{\alpha}}^{\mu}$ satisfies $g_{\mu\nu}e_{\hat{\alpha}}^{\mu}e_{\hat{\beta}}^{\nu}=\eta_{\hat{\alpha}\hat{\beta}}$, and $\eta_{\hat{\alpha}\hat{\beta}}$ is the Minkowski metric. At the point $P$, the local orthonormal base $\{e_{\hat{t}}, e_{\hat{r}}, e_{\hat{\theta}}, e_{\hat{\phi}} \} $ is related to the coordinate basis $\{\partial_t, \partial_r, \partial_{\theta}, \partial_{\phi} \} $ by
        \begin{equation}
            e_{\hat{\alpha}}=e_{\hat{\alpha}}^{\mu}\partial_{\mu}.
        \end{equation}
    As in \cite{2018PhRvD..97f4029W}, it is convenient to chose the expression of $e_{\hat{\alpha}}^{\mu}$ as
        \begin{equation}
            e_{\hat{\alpha}}^{\mu} =
            \left(
                \begin{array}{cccc}
                    \frac{1}{\sqrt{-g_{tt}}} & 0 & 0 & 0 \\
                    0 & \frac{1}{\sqrt{g_{rr}}} & 0 & 0 \\
                    0 & 0 & \frac{1}{\sqrt{g_{\theta\theta}}} & 0 \\
                    0 & 0 & 0 & \frac{1}{\sqrt{g_{\phi\phi}}} \\
                \end{array}
            \right).
        \end{equation}
    Therefore, in the local orthonormal frame at the point $P$, the non-zero components of magnetic field $B$ can be written as
        \begin{eqnarray}
            B^{\hat{r}}&=&B_{\hat{r}} = F_{\hat{\theta}\hat{\phi}} = -F_{\hat{\phi}\hat{\theta}} = \frac{4 m b r Q^{\frac{3}{2}} Z^{\frac{1}{2}} \cos\theta}{P^2 Y^2}, \\
            B^{\hat{\theta}}&=&B_{\hat{\theta}} = F_{\hat{\phi} \hat{r}} = - F_{\hat{r} \hat{\phi}} = \frac{2 m b Q^{\frac{3}{2}} \sin \theta \left(r^2+b^2 \cos^2\theta\right)}{P^2 Y^2}.
        \end{eqnarray}
    Similarly, in the local Cartesian P-frame, the energy-momentum tensor $T_{\hat{\alpha}\hat{\beta}}$ (\ref{emt0})-(\ref{emt1}) can be obtained from the tensor (\ref{emt0}) by a transformation $T_{\hat{\alpha}\hat{\beta}} = T_{\mu\nu} e_{\hat{\alpha}}^{\mu} e_{\hat{\beta}}^{\nu}$ as
    \begin{eqnarray}\label{emt0p}
        T_{\hat{0}\hat{0}} &=&-\frac{2 b^2 m^2 Q^3 \sin ^2\theta \left(\left(b^2 \cos ^2\theta+r^2\right)^2+4 r^2 Z \cot ^2\theta\right)}{P^4 Y^4},\\
        T_{\hat{1}\hat{1}} &=&-\frac{2 b^2 m^2 Q^3 \sin ^2\theta \left(\left(b^2 \cos ^2\theta+r^2\right)^2-4 r^2 Z \cot^2\theta\right)}{P^4 Y^4},\\
        T_{\hat{2}\hat{2}} &=&\frac{2 b^2 m^2 Q^3 \sin ^2\theta \left(\left(b^2 \cos ^2\theta+r^2\right)^2-4 r^2 Z \cot ^2\theta\right)}{P^4 Y^4},\\
        T_{\hat{1}\hat{2}} &=&T_{\hat{2}\hat{1}} =\frac{4 b^2 m^2 Q^3 r \sqrt{Z} \sin 2 \theta \left(b^2 \cos ^2\theta+r^2\right)}{P^4 Y^4},\\
        T_{\hat{3}\hat{3}} &=&-\frac{2 b^2 m^2 Q^3 \sin ^2\theta \left(\left(b^2 \cos ^2\theta+r^2\right)^2+4 r^2 Z \cot ^2\theta\right)}{P^4 Y^4}.
    \end{eqnarray}
    Since the accretion disk lies in the equatorial plane, the magnetic field distribution in the disk can be further simplified as
    \begin{equation}
        B^{\hat{r}} = 0, \quad \quad \quad B^{\hat{\theta}} = \frac{2 b m (r-m)^3}{r^4 (r-2 m)^2}
        \label{Bth}.
    \end{equation}
    This means that there is only the magnetic field in $\hat{\theta}$ direction in this disk and its distribution is related to the parameters $m$, $b$ and $r$. The corresponding energy-momentum tensor (\ref{emt0p}) becomes
    \begin{eqnarray}\label{emt0f}
        T_{\hat{0}\hat{0}}= T_{\hat{1}\hat{1}}= -T_{\hat{2}\hat{2}}= T_{\hat{3}\hat{3}}= -\frac{2 b^2 m^2 (r-m)^6}{r^8(r-2m)^4}.
    \end{eqnarray}
    We assume the emitter $P$ to move along a stable circular orbit in the equatorial plane around the black dihole with an angular velocity $\Omega$ in Eq.(\ref{elo}). Thus, the frequency of the emitter motion in the Bonnor dihole spacetime (\ref{metric}) can be written as
    \begin{equation}
       \nu_{P}=\frac{\Omega}{2\pi} = \frac{(r-2 m)^2}{2\pi r^2} \sqrt{\frac{2 m}{r(r^2-5 m r+6m^2)+2 b^2 m}}.
    \end{equation}
   Since the magnetic field direction $\hat{\theta}$ is perpendicular to the velocity direction $\hat{\phi}$, the magnetic field $B^{(\theta)}$ in the co-moving frame of the emitter $P$ can be expressed as
    \begin{equation}
        B^{(\theta)} = \gamma B^{\hat{\theta}},
    \end{equation}
    where $\gamma=1/ \sqrt{(1-\beta^2 / c^2)}$ is the Lorentz factor.  The energy-momentum tensor in the co-moving frame can be also obtained by the Lorentz transformation. For the sake of brevity, its form is not presented here.
    \begin{figure}[htb!]
        \includegraphics[width=16cm]{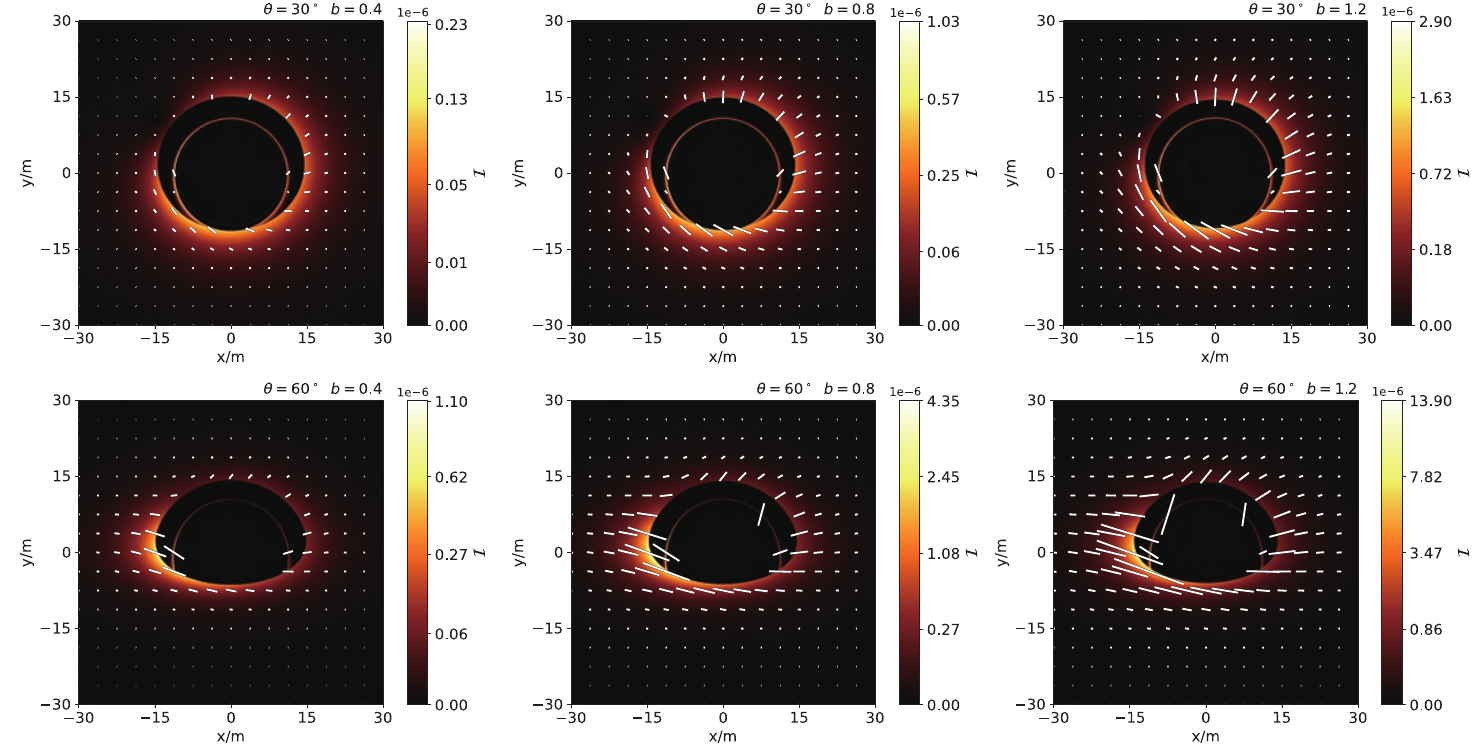}
        \caption{Polarized image of Bonnor black dihole arising from the synchrotron radiation with different $b$ and $\theta_0$. The observed linear polarization vectors are shown by the white lines. }
        \label{f6}
    \end{figure}
    \begin{figure}
        \includegraphics[width=12cm]{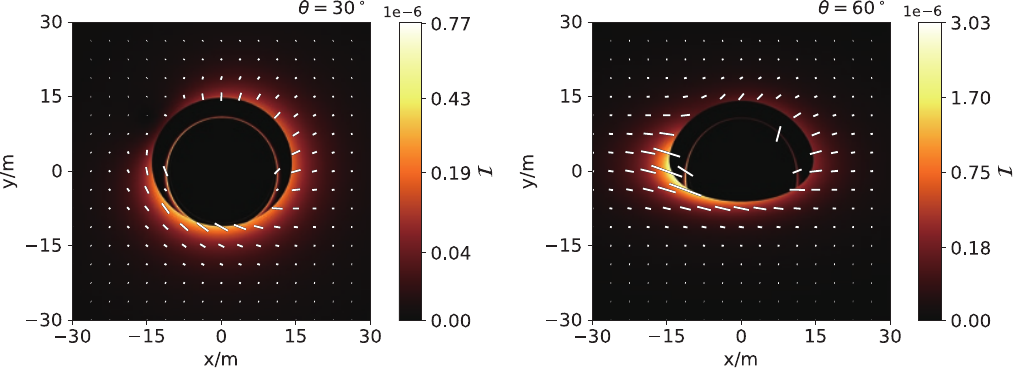}
        \caption{Polarized image of Schwarzschild black hole with the same magnetic field distribution as that of Bonnor black dihole with parameter $b=0.8$.}
        \label{f7}
    \end{figure}
    \begin{figure}
        \includegraphics[width=14cm]{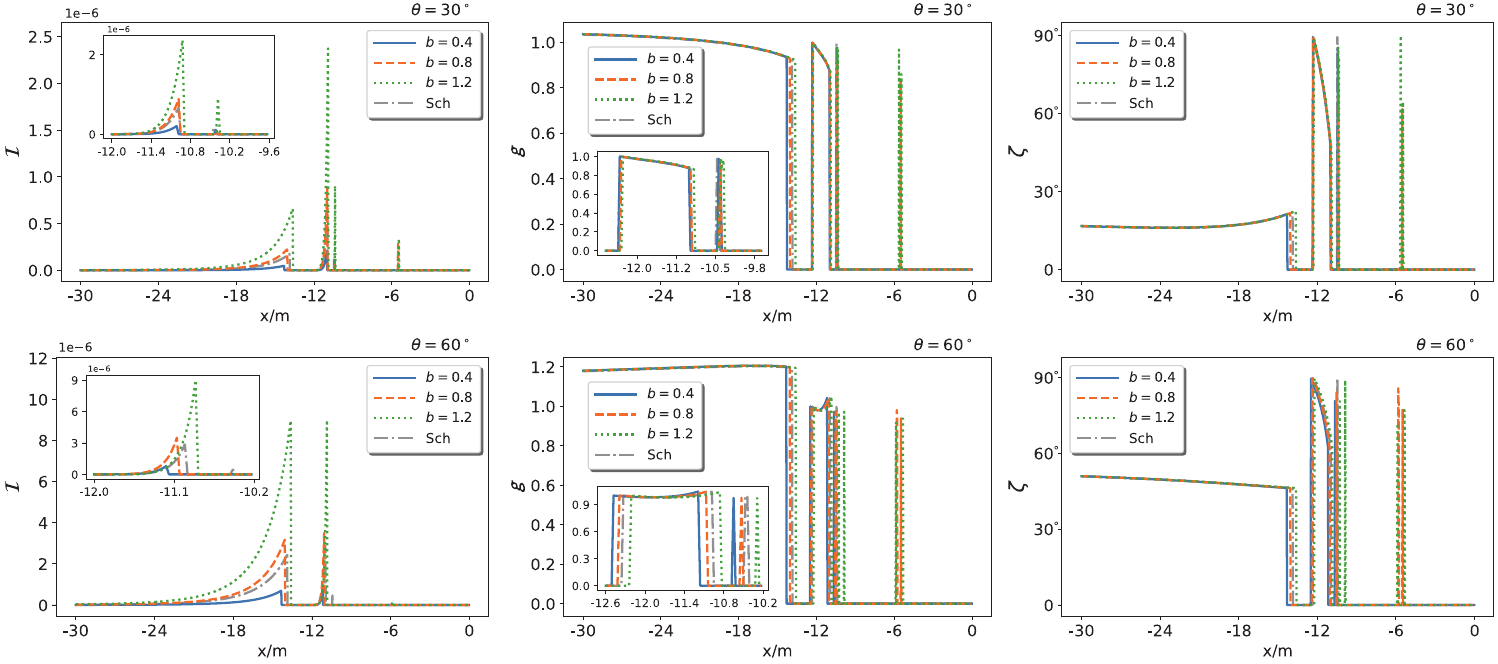}
        \caption{The distribution of the observed polarization intensity $\mathcal{I}$, the angle $\zeta$ and the redshift $g$ with $x$ along the line $y=0$ in Figs.\ref{f6} and \ref{f7}.}
        \label{f8}
    \end{figure}

    In the local frame co-moving with emitter, the three dimensional electric vector $\vec{E}$ of electromagnetic emission caused by synchrotron radiation is perpendicular to local photon three-momentum $\vec{p}=\left(p^{\left(r\right)},p^{\left(\theta\right)},p^{\left(\phi\right)}\right)$ and local magnetic field $\vec{B}=\left(B^{\left(r\right)},B^{\left(\theta\right)},B^{\left(\phi\right)}\right)$. Therefore, the spatial components of polarization vector $\vec{f}=\left(f^{\left(r\right)},f^{\left(\theta\right)},f^{\left(\phi\right)}\right) $ can be given by
    \begin{equation}
        \vec{f}=\frac{\vec{p} \times \vec{B}}{| \vec{p}| },
    \end{equation}
    and the corresponding temporal component is set to $f^{\left(t\right)}=0$.
    The polarization 4-vector also satisfies the normalization relationship as \cite{2021ApJ...912...35N,2021PhRvD.104d4060G}
    \begin{equation}
        f^{(\mu)}f_{(\mu)}=\sin^2\zeta| \vec{B}|^2,
    \end{equation}
    where $\zeta$ is the angle between photon momentum $\vec{p}$ and the magnetic field $\vec{B}$ and obeys
    \begin{equation}
        \sin \zeta = \frac{| \vec{p}\times \vec{B}| }{| \vec{p}|  | \vec{B} | }.
        \label{zeta}
    \end{equation}
   Making use of the polarization vector $f^{(\mu)}$ at the point $P$ in the co-moving frame, we can obtain the form of the polarization vector $f^{\mu}$ of the emitted photon at the point $P$ in the coordinate basis $\{\partial_t, \partial_r, \partial_{\theta}, \partial_{\phi} \} $ through some coordinate transformations. Furthermore, we can also obtain the polarization vector of photon reaching the observer from the emitter at the point $P$ by its parallel transport along the null geodesics of the photon in a spacetime \cite{2011MNRAS.410.1052S,2018ApJ...863....8P},
    \begin{equation}\label{ppf1}
        p^{\mu}\triangledown _{\mu}f^{\nu} = 0.
    \end{equation}
    Solving Eq.(\ref{ppf1}) together with the null geodesic equation of the photon, one can obtain the observed polarization vector at the observer $(f^{x}_{obs},\;f^{y}_{obs})$.
    \begin{figure}[htb!]
        \includegraphics[width=16cm]{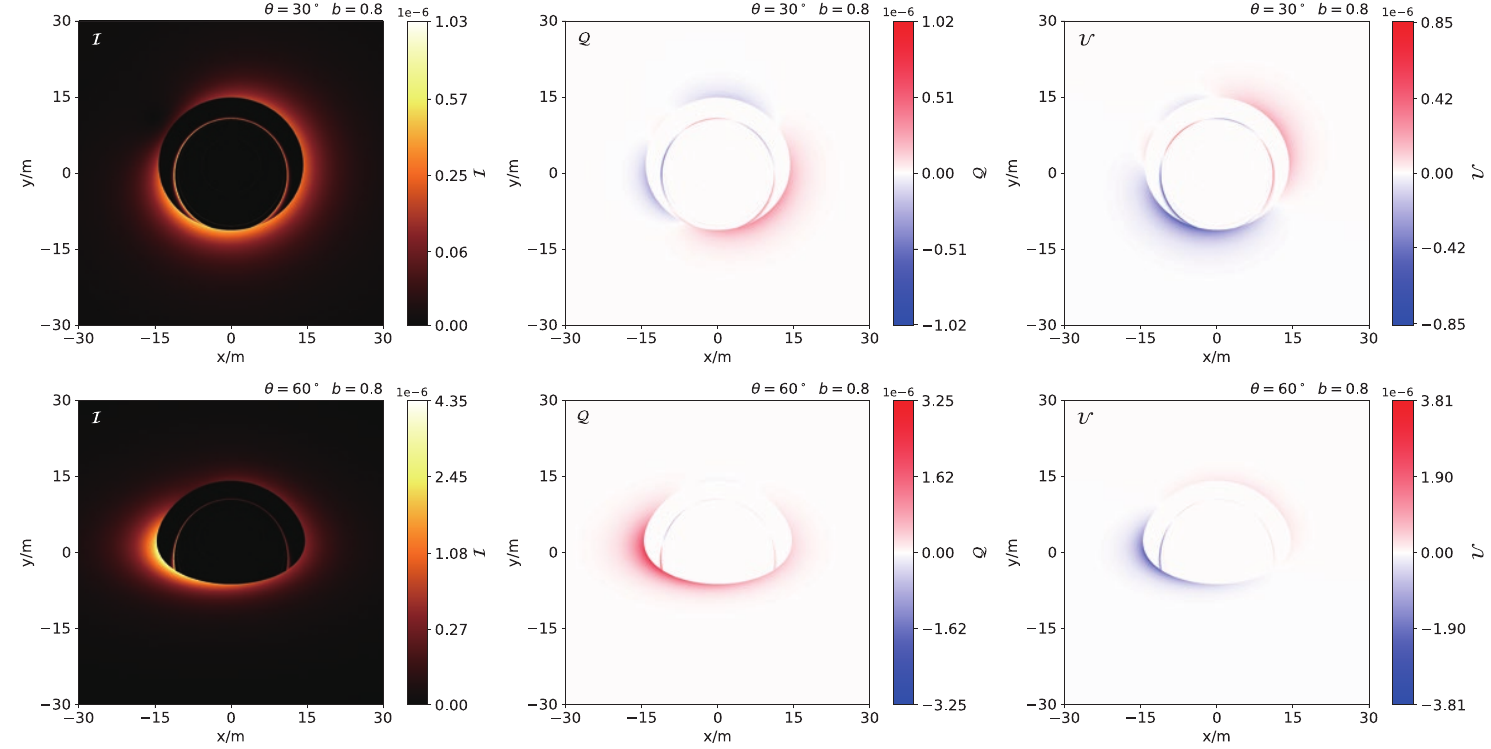}
        \caption{Distribution of the Stokes parameters $\mathcal{Q}$ and $\mathcal{U}$ in the image of Bonnor black dihole with parameter $b=0.8$. The panels in the top and bottom rows correspond to the cases with $\theta_0=30^{\circ}$ and $\theta_0=60^{\circ}$, respectively.}
        \label{f9}
    \end{figure}

    Following \cite{2021ApJ...912...35N}, the intensity $\mathcal{I}$ of linearly polarized light at the observer from the emitter can also be expressed as
    \begin{equation}\label{liangdutongbu}
        \mathcal{I}=l_p g^4 |\vec{B}|^2 \sin^2\zeta.
    \end{equation}
   Here $l_p$ is the geodesic path length for the photon traveling through the emitting region, and can be expressed as \cite{2021ApJ...912...35N}
    \begin{equation}
        l_p = \frac{p^{(t)}}{p^{(\theta)}} H.
    \end{equation}
    Here the disk height $H$ can be taken to be a constant for simplicity \cite{2021ApJ...912...35N,2021PhRvD.104d4060G}.   The redshift factor $g={\nu_o}/{\nu_s}$ includes the Doppler redshift caused by the motion of the emitter and gravitational redshift by the central black hole. Thus, the gravitational field of the central black hole has influence on the radiation intensity.
     Moreover, the intensity $\mathcal{I}$ and the Stokes parameters $\mathcal{Q}$ and $\mathcal{U}$ of the observed linearly polarization vector can be written as \cite{2021ApJ...912...35N,2021PhRvD.104d4060G}
    \begin{eqnarray}\label{QU00}
    \mathcal{I}=(f^{x}_{obs})^2+(f^{y}_{obs})^2,\quad\quad\quad
        \mathcal{Q}\equiv\mathcal{I}\cos\left(2\psi\right)=(f^{y}_{obs})^2-(f^{x}_{obs})^2,\quad\quad\quad
        \mathcal{U}\equiv\mathcal{I}\sin\left(2\psi\right)=-2f^{x}_{obs}f^{y}_{obs},
        \label{qu}
    \end{eqnarray}
    respectively. The angle $\psi$ is the so-called electric vector position angle (EVPA). With these quantities, we can analyze
    the polarization information in the image of Bonnor black dihole (\ref{metric}) with a thin accretion disk.
    \begin{figure}[htb!]
        \includegraphics[width=16cm]{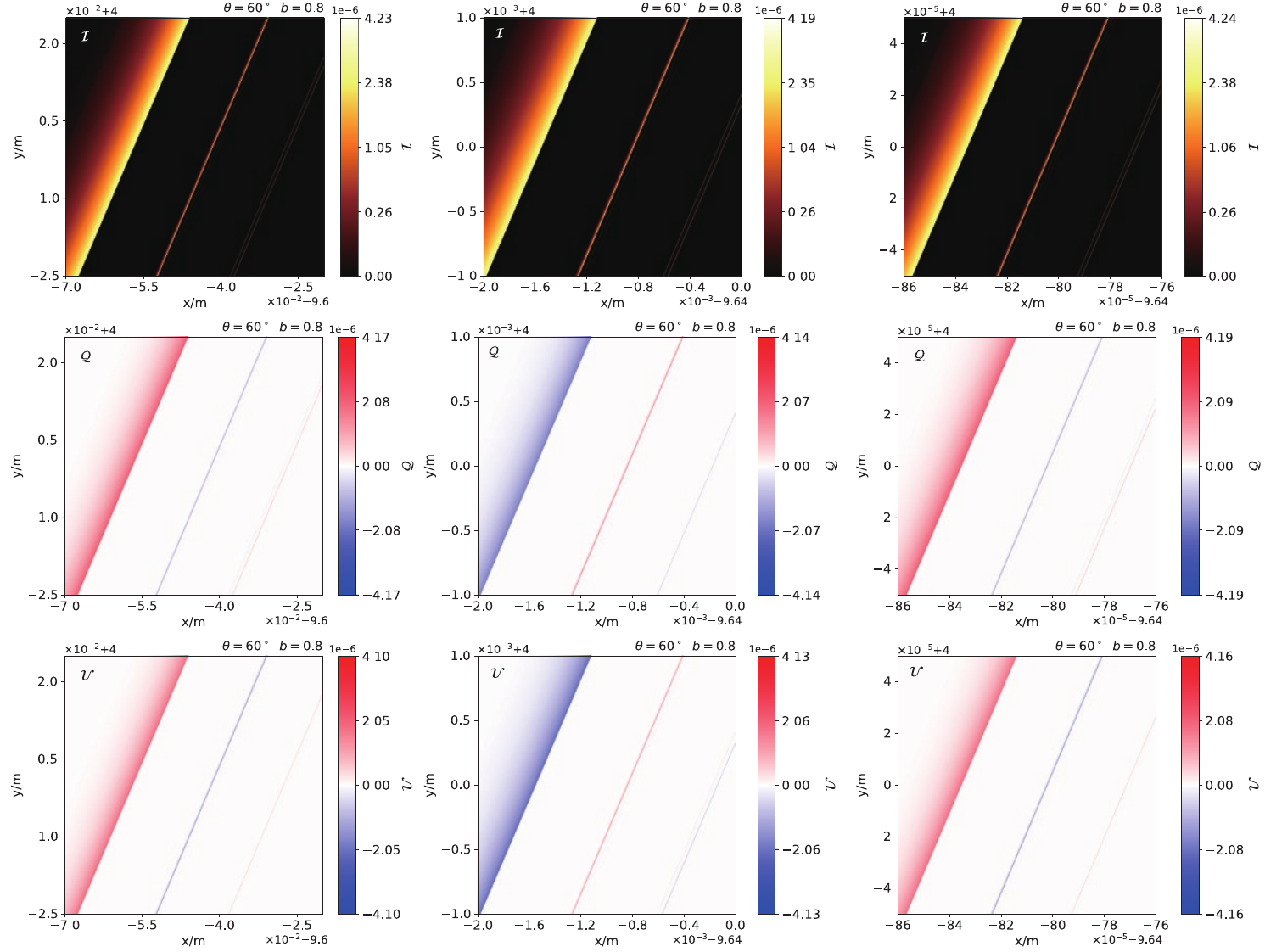}
        \caption{Enlarged image of $\mathcal{Q}$ and $\mathcal{U}$ near secondary image with $\theta_0=60^{\circ}$ and $b=0.8$. In each rows, the right panel is local magnification of the left panel.}
        \label{f10}
    \end{figure}

    Fig.\ref{f6} shows the polarized images of Bonnor black dihole arising from the synchrotron radiation with different magnetic parameters $b$ and inclination angles $\theta_0$. We find that the polarization intensity $\mathcal{I}$ of image decrease rapidly with the radius of the emitter in the disk, but increases with the magnetic parameter $b$. The main reason is that magnetic field $B^{\hat{\theta}}$ (\ref{Bth}) is a decreasing function of the radial coordinate $r$, but an increasing function of $b$. From Fig.\ref{f6}, one can also find that the effect of $b$ on the EVPA is tiny. Moreover, the dependence of Bonnor black dihole image caused by synchrotron radiation on the observation inclination $\theta_0$ is similar to that of the black dihole images by black body radiation.  Fig.\ref{f6} present the polarized images of Schwarzschild black hole with the same mass parameter and magnetic field distribution with $b=0.8$. We find that the distribution of the radiation around the Schwarzschild black hole is also similar to that around the Bonnor black dihole as in the case of black body radiation. Moreover, we also find that the maximum intensity in the Schwarzschild case is weaker than that in the Bonnor case.
    As in Fig.(\ref{f2}), we also present the polarization intensity distribution curve of image along the line $y=0$ in Fig.(\ref{f8}). It is shown that the intensity in the bright region of the image with the inclination angle $\theta_0=60^{\circ}$ is also stronger than that with $\theta_0=30^{\circ}$. Moreover, we note that the intensity of the secondary image is stronger than that of the direct image when $\theta_0=30^{\circ}$, which differs from that in the case $\theta_0=60^{\circ}$. It is understandable by a fact that the intensity $\mathcal{I}$ of image (\ref{liangdutongbu}) arising from synchrotron radiation depends on not only the angle $\zeta$, but also on the redshift factor $g$ in the propagation of photon from the source in the disk to the observer. From (\ref{f8}), the value of $\zeta$ corresponded to the direct image is less than that of the secondary image in the both cases $\theta_0=30^{\circ}$ and $\theta_0=60^{\circ}$. However, in the case with $\theta_0=30^{\circ}$, the redshift factor $g$ of the direct image is almost equivalent to that of the secondary one, which leads to that the intensity of the secondary image is stronger than that of the direct image when $\theta_0=30^{\circ}$. While, in the case with $\theta_0=60^{\circ}$, the redshift factor $g$ of the direct image is much larger than of the secondary one, and the total effect of the redshift factor $g$ together with the angle $\zeta$ leads to that the direct image is brighter than the secondary image. Comparing the intensity $\mathcal{I}$ of image, one can find that the black dihole image formed by synchrotron radiation is much brighter than that by black body radiation, but the intensity of black dihole image decreases more rapidly with the radial coordinate $r$ in the synchrotron radiation model.

    Fig.\ref{f9} shows the distribution of $\mathcal{Q}$ and $\mathcal{U}$ in the black dihole image with the parameter $b=0.8$ under different observation inclination angles. It is easy to see that the distribution of $\mathcal{Q}$ and $\mathcal{U}$ changes obviously with the change of observation inclination, which is consistent with that in Fig.\ref{f6}. However, the intensity distribution of $\mathcal{Q}$ is different from that of $\mathcal{U}$. The main reason is that $\mathcal{Q}$ and $\mathcal{U}$, from Eq.(\ref{QU00}), describe the intensity of linearly polarized light along two perpendicular directions, respectively. The polarization distribution on the secondary image can also be clearly seen from Fig.\ref{f9}.
    \begin{figure}
        \includegraphics[width=12cm]{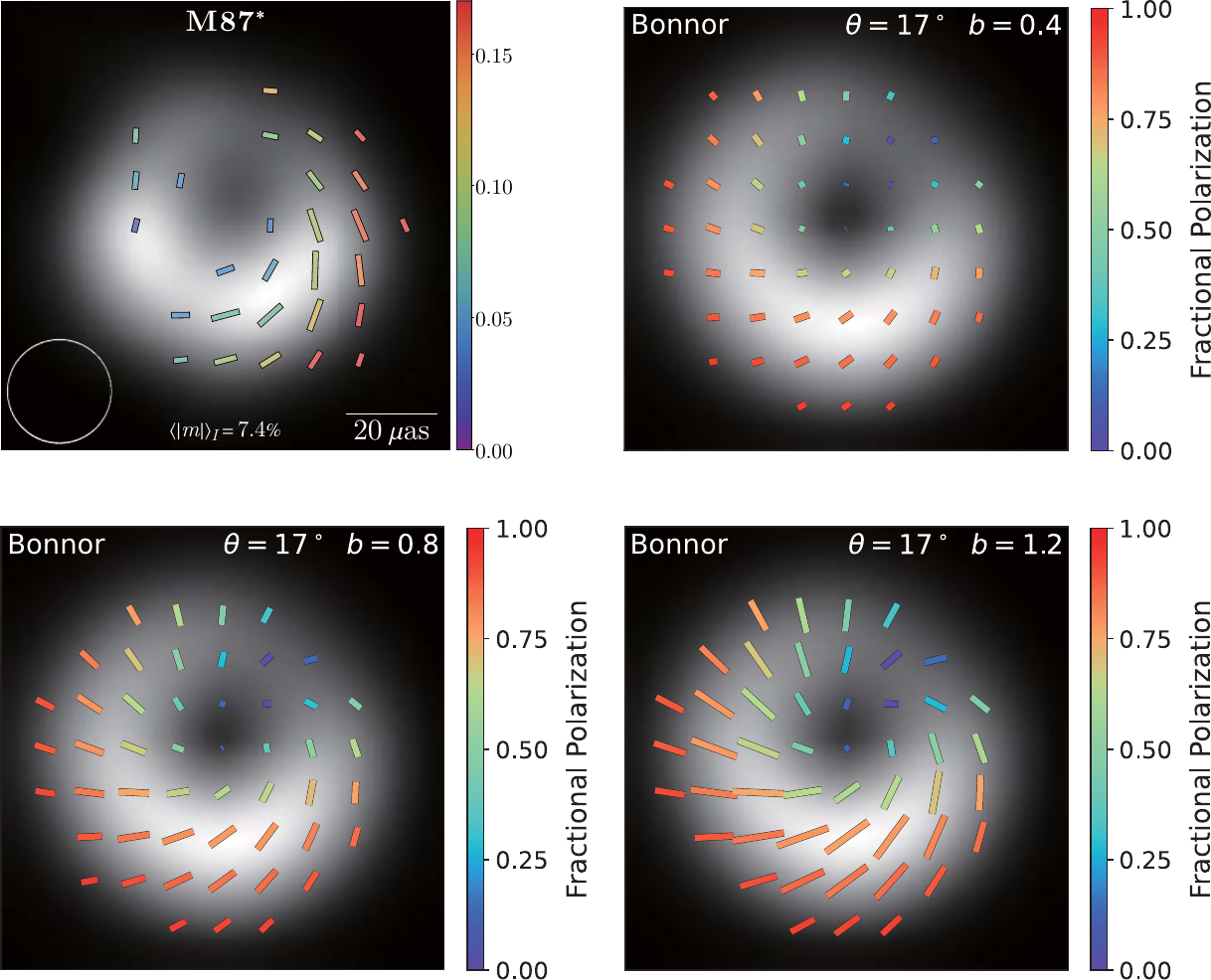}
        \caption{Comparison between the polarimetric image of Bonnor black dihole and the black hole M87* \cite{2021ApJ...912...35N}.}
        \label{f11}
    \end{figure}
    Zooming in the region near the secondary image, as shown in Fig.\ref{f10}, one can see the fine features similar to Fig.\ref{f5}. Moreover, we find that the signs of both $\mathcal{Q}$ and $\mathcal{U}$ are just the opposite for two adjacent indirect images. It could be caused by that the photons formed two adjacent indirect images are emitted from the upper and lower surfaces of accretion disk, respectively, which leads to the large difference of the corresponding polarization vectors.

    Finally, in Fig.\ref{f11}, we make a comparison of the polarimetric image between Bonnor black dihole and M87*. One can find that the polarimetric images of Bonnor black dihole for different $b$ own some spiral structures, which is similar to that of M87*. However, we also note that there exists significant difference between them. It implies that M87* may not belong to the family of Bonnor black diholes. This also confirms that the image of black hole depends on not only the black hole's properties itself, but also the matter around black hole and the corresponding radiation.

    \section{SUMMARY}
    \label{sec5}
    We have studied the image of Bonnor black dihole surrounded by a thin accretion disk. Here we considered two different radiation models in which the electromagnetic emission in the disk is dominated by black body radiation and synchrotron radiation, respectively. Our results show that the intensity of Bonnor black dihole image increases with the magnetic parameter $b$ and inclination angle $\theta_0$ in both models. However, the image of Bonnor black dihole in the synchrotron radiation model is one order of magnitude brighter than that in the black body radiation model. Moreover, the intensity of image decreases more rapidly with the radial coordinate $r$ in the synchrotron radiation model. We also present higher order images and some fine structures in the image of Bonnor black dihole surrounded by a thin accretion disk. For the black body radiation model, it is shown that the indirect images is dimmer than the direct images. However, for the synchrotron radiation model, we find that the intensity of the secondary image is stronger than that of the direct image when $\theta_0=30^{\circ}$, which differs from that in the case with black body radiation. Moreover, we also find that there exists an ``eyebrow shape" bright region in the black dihole shadows, which is caused by photon chaotic lensing. However, the pattern of ``eyebrow shape" bright region is not quite the same as that in \cite{2018PhRvD..97f4029W}, which is caused by the difference of light sources.

    We also present the polarization partners in the images of Bonnor black dihole in the synchrotron radiation case. The polarization partners and their distribution of $\mathcal{Q}$ and $\mathcal{U}$ depend sharply on the magnetic parameter $b$ and inclination angle $\theta_0$. For two adjacent indirect images, the signs of both $\mathcal{Q}$ and $\mathcal{U}$ are just the opposite. Finally, we make a comparison of the polarimetric image between Bonnor black dihole and M87*. Our result imply that M87* may not belong to the family of Bonnor black diholes. It also further confirms that the image of black hole depends on not only the black hole's properties itself, but also the matter around black hole and the corresponding radiation. In the real astrophysical situations, there exist enormous magnetic
    fields around the compacted objects with strong gravitational field.
    In the Milky Way, the discovered pulsar close to the galactic centre and the large Faraday rotation indicate that there is a dynamically relevant magnetic field about $30-100 G$ near the black hole Sgr A* \cite{2013Natur.501..391E}.  The EHT collaboration estimated the magnetic field strength in the plasma around M87* is $B\sim 1-30G$ \cite{2021ApJ...910L..13E}. Moreover, the observed radio pulsar neutron stars have strong magnetic fields typically of $10^{12} G$ \cite{Esposito:2020xll,Ioka:2000yb,Vasuth:2003yr}. However, at present, there is no evidence of the existence of Bonnor black dihole objects in nature. The study of Bonnor black dihole image and its polarization partners could provide some information to detect such kind of objects in the future.

    \section{\bf Acknowledgments}

    This work was supported by the National Natural Science Foundation of China under Grant No. 11875026 and 12035005.

    \bibliography{Bonnor-dihole}

\end{document}